\documentstyle[12pt]{article}

\begin{document}
\def\thebibliography#1{\section*{REFERENCES\markboth
 {REFERENCES}{REFERENCES}}\list
 {[\arabic{enumi}]}{\settowidth\labelwidth{[#1]}\leftmargin\labelwidth
 \advance\leftmargin\labelsep
 \usecounter{enumi}}
 \def\newblock{\hskip .11em plus .33em minus -.07em}
 \sloppy
 \sfcode`\.=1000\relax}
\let\endthebibliography=\endlist

\hoffset = -1truecm
\voffset = -2truecm

%%-Title and abstract page-%%
%%-move to normal A4-%%

\title{\large\bf
3D-4D Interlinkage Of qqq Wave Functions Under 3D Support For Pairwise 
Bethe-Salpeter Kernels 
}
\author{
{\normalsize\bf
A.N.Mitra \thanks{e.mail: (1) insa@giasdl01.vsnl.net.in(subj:a.n.mitra);
(2) anmitra@csec.ernet.in}
}\\
\normalsize 244 Tagore Park, Delhi-110009, India.}

\date{19 February 1998}

\maketitle

\begin{abstract}
Using the method of Green's functions within the framework of a 
Bethe-Salpeter formalism characterized by a pairwise $qq$ interaction 
with a Lorentz-covariant 3D support to its kernel, the 4D BS wave function 
for a system of three identical relativistic spinless quarks is 
reconstructed from the corresponding 3D quantities which satisfy a 
{\it fully connected} 3D BSE. This result is a 3-body generalization of a 
similar 3D-4D interconnection for the corresponding 2-body wave functions 
found earlier under identical conditions of a Lorentz-covariant 3D 
support to the corresponding BS kernel, (`CIA' for short), for the 
$q{\bar q}$ interaction. (The generalization from spinless to fermion 
quarks is straightforward). 
\par
	To set `CIA' for 3D Kernel support in the context of contemporary 
approaches to the $qqq$ baryon problem, a model scalar 4D $qqq$ BSE with 
pairwise contact interactions simulating the NJL-Faddeev equations, is 
worked out fully and compared with the `CIA' vertex function which reduces 
exactly to the 4D NJL-Faddeev form in the limit of zero spatial range. This 
consistency check is part of a detailed accounting of the CIA vertex structure 
whose physical motivation stems from the role of spectroscopy when considered 
as an integral part of any QCD-motivated dynamical investigation. \\\\

  
\end{abstract}

\newpage

%%-main body of paper-%%

%%-self numbering sections-%%
\section{Introduction: The Relativistic $qqq$ Problem }

The $qqq$ problem must obey the mathematical requirement of 
{\it connectedness} [1-4], signalled by the absence of $\delta$-functions
in its defining equation, Schroedinger (in 3D) or BSE (in 4D). Further, for 
a {\it relativistic} 2- or 3-body problem, the historical issue of 3D 
reduction from a 4D BSE has generated much interest from the outset [5-8] 
to more recent times [9-10], motivated by the Markov-Yukawa transversality
condition [11]. Physically, a 3D reduction of the BSE is linked to the 
problem of {\it observed} O(3)-like hadron spectra [12], while without
3D reduction, a 4D BSE in Wick-rotated form gives O(4)-like spectra [13].
In this paper we shall address both the issues of {\it connectedness} and
`3D-4D interlinkage' of the $qqq$ BS wave functions under 3D support for the 
pairwise kernels, termed `CIA' for short, as a 3-body extension of an 
earlier $q{\bar q}$ investigation [9].  
\par
	Our principal result is mathematical: Derivation of an explicit 
interconnection between the 3D and 4D forms of the $qqq$ BS wave functions, 
via Green's function techniques for the bound $qqq$ state. However, 
an apparent lack of familiarity with the 3D support ansatz, in the quantum 
mechanics literature, necessitates a prior exposition of its tenets vis-a-vis 
the more familiar ones, which we shall outline in the following sub-sections, 
keeping in mind the overall requirement of connectedness in a 3-particle 
amplitude [1-4], signalled by the {\it {absence of any delta function}} in 
its structure, either explicitly or through its defining equation. Originally 
derived within a 3D framework [1-4] whose prototype dynamics is the 
Schroedinger equation, its basic logic applies to the 4D BSE framework 
with pairwise kernels, with or without 3D reduction.  

\subsection{3D Reduction of 4D BSE}

The problem of 3D reduction of the BSE itself has had a long history [5-8],
born out of certain intuitive compulsions to keep to a common time for the
internal components of a bound state composite. These have ranged from the 
instantaneous approximation [5] through the quasi-potential approach [6], 
to other variants [7,8] in which the interaction of the constituents can be 
given a physical meaning in their respective mass shells, a result that can 
also be justified within the tenets of {\it local} field theory [8]. In all 
these methods the starting BSE is formally 4D in all details, {\it {including 
its kernel}}, but the associated propagators are manipulated in various subtle
ways ranging from an `overall instantaneity' (the quisi-potential approach
[6]) to their individual `on-shellness' in varying degrees [7,8]. In all 
these approaches, a common feature is that once the initial BSE is thus
reduced to a 3D form, there is no getting back to its original 4D form.
\par    
	An alternative approach which is of more recent origin [9,10], is
based on the Markov-Yukawa Transversality Condition [11] wherein a 3D support 
is postulated at the outset to the {\it kernel} of an otherwise 4D BSE,  
albeit in a Lorentz-covariant form, but the propagators are left untouched
in their original 4D form. This is achieved by simply demanding that the 
kernel $K$ of the pairwise interaction be a function of only the component 
${\hat q}$ of the relative 4-momentum $q$  which is {\it transverse} to
the total 4-momentum $P$ of the composite, viz., 
${\hat q}_\mu = q-q.PP_\mu/P^2$, so that ${\hat q}.P = 0$ identically. 
This may be regarded as a sort of complementary strategy to the approaches 
[6-8] wherein the propagators are manipulated but the kernel is left 
untouched. The main difference between the two approaches is that while
in [6-8] the original 4D BSE must be given up for good in favour of the 
reduced 3D BSE as a fresh staring point of the dynamics, the alternative 
approach [9,10] based on the Transversality Condition [11] allows both forms 
to be used interchangeably according to the demands of physics. For, not only 
does this ansatz of `Covariant Instaneity' (termed CIA for short [9]) lead 
to an exact 3D reduction of the BSE, but also facilitates an equally exact 
reconstruction of the original 4D BSE form without extra charge [9], so that 
both forms are completely equivalent. For a 2-quark problem this was shown 
explicitly some time ago [9] under `CIA' which gives a concrete realization of a 
`two-tier' approach [14] wherein  the 3D reduction of the original 4D BSE 
serves for the dynamics of the observed O(3)-like spectra [12],  while 
the reconstructed 4D BS wave function  provides a natural language for 
applications to various transition amplitudes as 4D quark loop integrals 
[14,9]. The exact interconnection that CIA yields between the 3D and 4D 
forms of the BSE shows up through the expression of the 4D BS vertex function 
${\Gamma}$  as a simple product of only 3D quantities, viz., $D \times \phi$, 
where $D$ and ${\phi}$ are the 3D denominator and wave functions respectively, 
satisfying a relativistic Schroedinger-like equation [9]. We next turn to the
physical basis of this method [15], termed 3D-4D-BSE in the following.     

\subsection{3D-4D BSE vis-a-vis other QCD Motivated Methods}

The  physical basis of  3D-4D-BSE [15] vis-a-vis other 4D BSE-cum-SDE 
approaches [16], is  dynamical breaking of chiral symmetry, or $DB{\chi}S$ 
for short, a la NJL [17]. While the original NJL model [17] was conceived as a 
contact interaction, its vital feature of chiral symmetry and its breaking, 
which it shares with other effective theories [18,19] nevertheless is a key 
ingredient which various workers have attempted to bring out [20-23] from 
QCD premises in the low frequency limit to simulate its non-perturbative 
features. Of particular interest to the present investigation is a non-local 
4-quark interaction mediated by vector exchange [21], as a prototype of the 
non-perturbative gluon propagator, which offers a generalized $DB{\chi}S$ 
mechanism for generating a momentum-dependent mass-function $m(p)$ via
the Schwinger-Dyson equation (SDE). As is well known, this mechanism 
accounts for the bulk of the `constituent' mass of $ud$ quarks via Politzer 
additivity [24]. This formalism leads to the conventional 4D BSE-SDE type 
framework [21,16] which was discussed in [15] vis-a-vis 3D-4D-BSE [9]. 
\par
	A related aspect of QCD concerns its structure in the large $N_C$ 
limit, when it reduces to an effective theory of weakly interacting mesons 
and glueballs [25], while baryons emerge as solitonic  configurations 
in the background meson fields [26].  These ideas were later concretized by 
many authors [22,23] using functional integration techniques [27], which 
combined both the  $DB{\chi}S$ and large $N_C$ aspects of QCD in a systematic 
reduction procedure to give rise to an effective action  involving local
meson fields. While the large $N_C$ limit motivates an effective 4-quark 
interaction of the non-local NJL-type [21], the vacuum degeneracy of the 
effective action so derived [22,23] exhibits a complex structure arising 
from $DB{\chi}S$, viz., after integrating out over the quark d.o.f.'s, the 
effective action of the chiral $\pi, \sigma$- field (treated classically) 
is approximately a sum of bosonic and fermionic parts, of which the former 
gives rise to mesonic excitations [28], while the latter provides the 
solitonic solution corresponding to `baryon-number-one'[29].        
\par
	Despite a common QCD basis for the  4D BSE-SDE [16] and the 
solitonic [26] pictures, their actual technologies are so different
that only the former [16], and not the latter [26], is directly relevant to 
our discussion. In particular, the Lagrangian formulation [21] 
to which our 3D-4D BSE comes closest, corresponds to an  effective 
$q{\bar q}/qq$ interaction with a gluon-like propagator to simulate both the 
perturbative and non-perturbative regimes, as originally envisaged [30]
in the instantaneous approximation [5]. Subsequently this kernel was given 
a covariant 3D basis [9] in the spirit of [11], and the reduced 3D BSE found 
to agree with the spectra of mesons [31], as well as of baryons [32], 
under a common parametrization for the gluon propagator with 2 basic 
constants plus the constituent quark mass $m_q \approx 270 MeV$. This last
turned out to be consistent with the chiral symmetry breaking SDE solution 
in the low momentum limit [15], so that this quantity is no longer a free 
parameter. Thus this BSE-cun-SDE approach [15] with an effective gluon-like 
propagator [30] produces the usual $DB{\chi}S$ scenario, in harmony with the 
standard 4D BSE-SDE picture [21,16], except for its hybrid 3D-4D structure, 
designed to unify the 3D hadron spectra [31-32] with the 4D quark-loop 
integrals [9,15]. With this physical background for 3D-4D BSE,  we now turn
to the $qqq$ sector.    
    
\subsection{ The qqq BSE Structure under 3D Kernel Support}

Does an interconnection similar to the $q{\bar q}$ case [9] exist in the 
corresponding BS amplitudes for a $qqq$ system under identical conditions of 
3D support to the pairwise BS kernel ?  This question is of great practical 
value  since the 3D reduction of the 4D BSE  already produces {\it {fully 
connected}} integral equations for both the $q{\bar q}$ [31] and $qqq$ [32] 
systems with explicit solutions for the 3D wave functions. Therefore a 
reconstruction [33] of the 4D $qqq$ wave (vertex) function in terms of the 
corresponding 3D quantities is a vital ingredient for applications to various 
types of {\it {transition amplitudes}} involving $qqq$ baryons, analogously 
to the ${\bar q}q$ case [9,15]. To keep the technicalities to the barest 
minimum, we shall consider 3 identical spinless particles for simplicity 
and definiteness, which however need not detract from the generality of 
the ensuing singularity structures. The answer is found to be in the 
affirmative, except for the recognition that a 3D support to the pairwise 
BS kernel implies a truncation of the Hilbert space. Such truncation, 
while still allowing an unambiguous reduction of the BSE from the 4D to the 
3D level, nevertheless leaves an information gap in the 
{\it {reverse direction}}, viz., from 3D to 4D [33] for any $n$-body system  
except for $n = 2$ where both transitions are exactly reversible (a sort of 
degenerate situation) [9]. The extra charge needed to complete the reverse 
transition comes in the form of a 1D delta-function which however has nothing 
to do with connectedness [3,4] of an $n$-body amplitude, as will be formally 
demonstrated in Sects.2-4, and fully accounted for in Sect.5 through a 
detailed comparison with a model NJL-Faddeev problem which has proved popular 
in recent years [34]. (Other methods include i) non-topological solitons 
[29,35]; and ii) QCD bosonization with diquarks [36], but these will not 
be considered).
\par  
	Regarding 3D-4D-BSE [33]  versus 4D NJL-Faddeev [34], the following 
perspective is in order: The 3D-4D-BSE pairwise kernel with covariant 
instantaneity (CIA) has a contact interaction only in the 1D time-like 
variable, while the  corresponding 4D-NJL kernel exhibits contact behaviour 
in all the 4D space-time variables. The consistency of the two approaches is 
ensured by checking that the latter is the (expected) $K = const$ limit of 
the former, and further that the Hilbert space information gap in the way of 
reconstruction of the 4D BS wave function in terms of 3D quantities [33] 
vanishes in this limit, so that the full 4D structure of NJL-Faddeev [34] is 
automatically recovered. 
\par
	The paper which makes use of Green's function techniques, is 
organized as follows. In Sec.2 we rederive the 3D-4D interconnection [9] 
for a two-body system, now by the Green's function method, whence we 
reproduce the previously derived result [9] for the corresponding BS wave 
functions in 3D and 4D forms. In Sec.3, the BSE for the 4D Green's function 
for 3 identical spinless quarks ($q$), is reduced to the 3D form by 
integrating w.r.t. {\it two} internal time-like momenta, and in so doing, 
introducing 3D Green's functions satisfying a fully connected 3D BSE, 
free from $\delta$-functions, as anticipated from BS wave function 
studies [32,37]. With this 3D BSE as the check point, Sec.4 gives a 
reconstruction of the full 4D Green's function in terms of its (partial) 
3D counterparts, so as to satisfy exactly the above 3D BSE, {\it after} 
integration w.r.t. the relevant time-like momenta. The resultant 
vertex function, eq.(4.10), has a 1D $\delta$-function singularity which 
admits a (Fermi-like) `pseudopotential' interpretation [38]. 
\par
	Sec.5 is devoted to the solution of a 4D NJL-Faddeev problem with 
scalar-isoscalar quarks in pairwise contact interaction, leading to a 4D 
baryon-quark-diquark vertex function (5.9), and subjected to a pointwise 
comparison with (4.10) [sec.(5.4)], and its structure accounted for. Sec.6 
sums up our main results as obtained in Secs.3-4, vis-a-vis the corresponding 
results of NJL-Faddeev [34] as obtained in Sec.5, and briefly indicates the 
technical issues arising from the inclusion of spin. It concludes with a 
brief comparison with some contemporary approaches, [6-8,16,34-36], especially
in relation to their respective roles in treating the spectroscopy sector as 
an integral part of the $qqq$ dynamics.

%%-self numbering subsections-%%
\section{3D-4D Interconnection For ${\bar q}q$ System}
	
	If the BSE for a spinless ${\bar q}q$ system has a 3D support for 
its kernel $K$ as $K({\hat q,\hat q'})$ where ${\hat q}$ is the 
component of the relative momentum $q = (p_1- p_2)/2$ {\it orthogonal} to 
the total hadron 4-momentum $P = p_1 + p_2$, then the 4D hadron-quark 
vertex function $\Gamma$ is a function of ${\hat q}$ only [9]. For this 
2-body case the 4D and 3D forms of the BSE are exactly reversible without 
further assumptions. For the 3-body case a corresponding 3D-4D connection 
was obtained on the basis of semi-intuitive arguments [37] whose formal 
derivation is the central aim of this paper. To that end we shall formulate 
the 4D and 3D BSE's in terms of Green's functions to derive the 3D-4D 
connection for a two-body system in terms of their respective Green's 
functions, in preparation for the generalization to the three-body 
case in the next two sections.
\par
	We shall mostly use the notation and phase conventions of [8,12] for
the various quantities (momenta, propagators, etc). The 4D $qq$ Green's
function $G(p_1p_2 ; {p_1}'{p_2}')$ near a {\it bound} state satisfies a 
4D BSE without the inhomogeneous term, viz. [9,37],
\setcounter{equation}{0}
\renewcommand{\theequation}{2.\arabic{equation}}
\begin{equation}
i(2\pi)^4 G(p_1 p_2;{p_1}'{p_2}') = {\Delta_1}^{-1} {\Delta_2}^{-1} \int
d{p_1}'' d{p_2}'' K(p_1 p_2;{p_1}''{p_2}'') G({p_1}''{p_2}'';{p_1}'{p_2}')    
\end{equation}
where
\begin{equation}
\Delta_1 = {p_1}^2 + {m_q}^2 , 
\end{equation}
and $m_q$ is the mass of each quark. Now using the relative 4- momentum 
$q = (p_1-p_2)/2$ and total 4-momentum $P = p_1 + p_2$ 
(similarly for the other sets), and removing a $\delta$-function
for overall 4-momentum conservation, from each of the $G$- and $K$- 
functions, eq.(2.1) reduces to the simpler form    
\begin{equation}
i(2\pi)^4 G(q.q') = {\Delta_1}^{-1} {\Delta_2}^{-1}  \int d{\hat q}'' 
Md{\sigma}'' K({\hat q},{\hat q''}) G(q'',q')
\end{equation}
where ${\hat q}_{\mu} = q_{\mu} - {\sigma} P_{\mu}$, with 
$\sigma = (q.P)/P^2$, is effectively 3D in content (being orthogonal to
$P_{\mu}$). Here we have incorporated the ansatz of a 3D support for the
kernel $K$ (independent of $\sigma$ and ${\sigma}'$), and broken up the 
4D measure $dq''$ arising from (2.1) into the product 
$d{\hat q}''Md{\sigma}''$ of a 3D and a 1D measure respectively. We have 
also suppressed the 4-momentum $P_{\mu}$ label, with $(P^2 = -M^2)$, in 
the notation for $G(q.q')$.
\par
	Now define a fully 3D Green's function ${\hat G}({\hat q},{\hat q}')$
as [9,37] 
\begin{equation}
{\hat G}({\hat q},{\hat q}') = \int \int M^2 d{\sigma}d{\sigma}'G(q,q')
\end{equation}
and two (hybrid) 3D-4D Green's functions ${\tilde G}({\hat q},q')$,
${\tilde G}(q,{\hat q}')$ as
\begin{equation}
{\tilde G}({\hat q},q') = \int Md{\sigma} G(q,q');
{\tilde G}(q,{\hat q}') = \int Md{\sigma}' G(q,q');
\end{equation} 
Next, use (2.5) in (2.3) to give    
\begin{equation}
i(2\pi)^4 {\tilde G}(q,{\hat q}') = {\Delta_1}^{-1} {\Delta_2}^{-1} 
\int dq'' K({\hat q},{\hat q}''){\tilde G}(q'',{\hat q}')  
\end{equation}
Now integrate both sides of (2.3) w.r.t. $Md{\sigma}$ and use the result [9]
\begin{equation}
\int Md{\sigma}{\Delta_1}^{-1} {\Delta_2}^{-1} = 2{\pi}i D^{-1}({\hat q});
\quad D({\hat q}) = 4{\hat \omega}({\hat \omega}^2 - M^2/4);\quad 
{\hat \omega}^2 = {m_q}^2 + {\hat q}^2 
\end{equation}
to give a 3D BSE w.r.t. the variable ${\hat q}$, while keeping the other 
variable $q'$ in a 4D form:
\begin{equation}  
(2\pi)^3 {\tilde G}({\hat q},q') = D^{-1} \int d{\hat q}''  
K({\hat q},{\hat q}'') {\tilde G}({\hat q}'',q')
\end{equation}
Now a comparison of (2.3) with (2.8) gives the desired connection between 
the full 4D $G$-function and the hybrid ${\tilde G({\hat q}, q')}$-function: 
\begin{equation}  
2{\pi}i G(q,q') = D({\hat q}){\Delta_1}^{-1}{\Delta_2}^{-1}
{\tilde G}({\hat q},q')
\end{equation}
which is the Green's function counterpart, {\it{near the bound state}}, 
of the same result [9] connecting the corresponding BS wave functions.
Again, the symmetry of the left hand side of (2.9) w.r.t. $q$ and $q'$ 
allows us to write the right hand side with the roles of $q$ and $q'$ 
interchanged. This gives the dual form   
\begin{equation}  
2{\pi}i G(q,q') = D({\hat q}'){{\Delta_1}'}^{-1}{{\Delta_2}'}^{-1}
{\tilde G}(q,{\hat q}')
\end{equation}
which on integrating both sides w.r.t. $M d{\sigma}$ gives
\begin{equation}  
2{\pi}i{\tilde G}({\hat q},q') = D({\hat q}'){{\Delta_1}'}^{-1}
{{\Delta_2}'}^{-1}{\hat G}({\hat q},{\hat q}'). 
\end{equation}
Substitution of (2.11) in (2.9) then gives the symmetrical form
\begin{equation}  
(2{\pi}i)^2 G(q,q') = D({\hat q}){\Delta_1}^{-1}{\Delta_2}^{-1}
{\hat G}({\hat q},{\hat q}')D({\hat q}'){{\Delta_1}'}^{-1}
{{\Delta_2}'}^{-1}
\end{equation}
Finally, integrating both sides of (2.8) w.r.t. $M d{\sigma}'$, we 
obtain a fully reduced 3D BSE for the 3D Green's function:
\begin{equation}  
(2\pi)^3 {\hat G}({\hat q},{\hat q}') = D^{-1}({\hat q} \int d{\hat q}''
K({\hat q},{\hat q}'') {\hat G}({\hat q}'',{\hat q}')
\end{equation}
Eq.(2.12) which is valid near the bound state pole (since the 
inhomogeneous term has been dropped for simplicity) expresses the desired 
connection between the 3D and 4D forms of the Green's functions; and 
eq(2.13) is the determining equation for the 3D form. A spectral analysis 
can now be made for either of the 3D or 4D Green's functions in the 
standard manner, viz., 
\begin{equation}  
G(q,q') = \sum_n {\Phi}_n(q;P){\Phi}_n^*(q';P)/(P^2 + M^2) 
\end{equation}
where $\Phi$ is the 4D BS wave function. A similar expansion holds for 
the 3D $G$-function ${\hat G}$ in terms of ${\phi}_n({\hat q})$. Substituting
these expansions in (2.12), one immediately sees the connection between 
the 3D and 4D wave functions in the form:
\begin{equation}  
2{\pi}i{\Phi}(q,P) = {\Delta_1}^{-1}{\Delta_2}^{-1}D(\hat q){\phi}(\hat q)
\end{equation}
whence the BS vertex function becomes $\Gamma = D \times \phi/(2{\pi}i)$
as found in [9]. We shall make free use of these results, taken as $qq$ 
subsystems, for our study of the $qqq$ $G$-functions in Sections 3 and 4.  

\section{qqq Green's Function: 3D Reduction of BSE}

	As in the two-body case, and in an obvious notation for various 
4-momenta (without the Greek suffixes), we consider the most general 
Green's function $G(p_1 p_2 p_3;{p_1}' {p_2}' {p_3}')$ for 3-quark 
scattering {\it near the bound state pole} (for simplicity) which allows       
us to drop the various inhomogeneous terms from the beginning. Again we 
take out an overall delta function $\delta(p_1 + p_2 + p_3 - P)$ from the
$G$-function  and work with {\it two} internal 4-momenta for each of the 
initial and final states defined as follows [37]:

\setcounter{equation}{0}
\renewcommand{\theequation}{3.\arabic{equation}}

\begin{equation}  
{\sqrt 3}{\xi}_3 =p_1 - p_2 \ ; \quad  3{\eta}_3 = - 2p_3 + p_1 +p_2
\end{equation}
\begin{equation}  
P = p_1 + p_2 + p_3 = {p_1}' + {p_2}' + {p_3}'
\end{equation}
and two other sets ${\xi}_1,{\eta}_1$ and ${\xi}_2,{\eta}_2$ defined by 
cyclic permutations from (3.1). Further, as we shall be considering pairwise
kernels with 3D support, we define the effectively 3D momenta ${\hat p}_i$, 
as well as the three (cyclic) sets of internal momenta 
${\hat \xi}_i,{\hat \eta}_i$, (i = 1,2,3) by [37]:
\begin{equation}
{\hat p}_i = p_i - {\nu}_i P \ ;\quad  {\hat {\xi}}_i = {\xi}_i - s_i P\  ;
\quad
{\hat {\eta}}_i = {\eta}_i - t_i P 
\end{equation}
\begin{equation}  
{\nu}_i = (P.p_i)/P^2\  ;\quad s_i = (P.\xi_i)/P^2 \ ;\quad t_i = 
(P.\eta_i)/P^2 \end{equation}
\begin{equation}  
{\sqrt 3} s_3 = \nu_1 - \nu_2 \ ;\quad 3 t_3 = -2 \nu_3 + \nu_1 + \nu_2 \ 
\ ( + {\rm cyclic permutations})
\end{equation}

The space-like momenta ${\hat p}_i$ and the time-like ones $\nu_i$ 
satisfy [37] 
\begin{equation}  
{\hat p}_1 + {\hat p}_2 + {\hat p}_3 = 0\  ;\quad \nu_1 + \nu_2 + \nu_3 = 1
\end{equation}
Strictly speaking, in the spirit of covariant instantaneity, we should 
have taken the relative 3D momenta ${\hat \xi},{\hat \eta}$ to be in the 
instantaneous frames of the concerned pairs, i.e., w.r.t. the rest frames
of $P_{ij} = p_i +p_j$; however the difference between the rest frames of 
$P$ and $P_{ij}$  is small and calculable [37], while the use of a common 
3-body rest frame $(P = 0)$ lends considerable simplicity and elegance to 
the formalism.   
\par
	We may now use the foregoing considerations to write down the BSE 
for the 6-point Green's function in terms of relative momenta, on closely 
parallel lines to the 2-body case. To that end note that the 2-body 
relative momenta are $q_{ij} = (p_i - p_j)/2 = {\sqrt 3}{\xi_k}/2$, where 
(ijk) are cyclic permutations of (123). Then for the reduced $qqq$ Green's
function, when the {\it last} interaction was in the (ij) pair, we may use 
the notation $G(\xi_k \eta_k ; {\xi_k}' {\eta_k}')$, together with 'hat' 
notations on these 4-momenta when the corresponding time-like components 
are integrated out. Further, since the pair $\xi_k,\eta_k$ is 
{\it {permutation invariant}} as a whole, we may choose to drop the index 
notation from the complete $G$-function to emphasize this symmetry as and 
when needed. The $G$-function for the $qqq$ system satisfies, in the 
neighbourhood of the bound state pole, the following (homogeneous) 4D BSE
for pairwise $qq$ kernels with 3D support:
\begin{equation}  
i(2\pi)^4 G(\xi \eta ;{\xi}' {\eta}') = \sum_{123}
{\Delta_1}^{-1} {\Delta_2}^{-1} \int d{{\hat q}_{12}}'' M d{\sigma_{12}}''
K({\hat q}_{12}, {{\hat q}_{12}}'') G({\xi_3}'' {\eta_3}'';{\xi_3}' {\eta_3}')
\end{equation}
where we have employed a mixed notation ($q_{12}$ versus $\xi_3$) to stress
the two-body nature of the interaction with one spectator at a time, in a 
normalization directly comparable with eq.(2.3) for the corresponding 
two-body problem. Note also the connections 
\begin{equation}  
\sigma_{12} = {\sqrt 3}{s_3}/2   ;\quad 
{\hat q}_{12} = {\sqrt 3}{{\hat \xi}_3}/2  ; \quad {\hat \eta}_3 = 
-{\hat p}_3, \quad etc 
\end{equation}  
The next task is to reduce the 4D BSE (3.7) to a fully 3D form through a 
sequence of integrations w.r.t. the time-like momenta $s_i,t_i$ applied 
to the different terms on the right hand side, {\it {provided both}} 
variables are simultaneously permuted. We now define the following fully 
3D as well as mixed (hybrid) 3D-4D $G$-functions according as one or more 
of the time-like $\xi,\eta$ variables are integrated out:
\begin{equation}  
{\hat G}({\hat \xi} {\hat \eta};{\hat \xi}' {\hat \eta}') = 
\int \int \int \int ds dt ds' dt' G(\xi \eta ; {\xi}' {\eta}')  
\end{equation}  
which is $S_3$-symmetric.
\begin{equation}  
{\tilde G}_{3\eta}(\xi {\hat \eta};{\xi}' {\hat \eta}') = 
\int \int dt_3 d{t_3}' G(\xi \eta ; {\xi}' {\eta}');
\end{equation}  
\begin{equation}  
{\tilde G}_{3\xi}({\hat \xi}  \eta;{\hat \xi}' {\eta}') = 
\int \int ds_3 d{s_3}' G(\xi \eta ; {\xi}' {\eta}');
\end{equation} 
The last two equations are however {\it not} symmetric w.r.t. the 
permutation group $S_3$, since both the variables ${\xi,\eta}$ are not 
simultaneously transformed; this fact has been indicated in eqs.(10,11) 
by the suffix ``3" on the corresponding (hybrid) ${\tilde G}$-functions,
to emphasize that the `asymmetry' is w.r.t. the index ``3". We shall term 
such quantities ``$S_3$-indexed", to distinguish them from $S_3$-symmetric 
quantities as in eq.(3.9). The full 3D BSE for the ${\hat G}$-function is 
obtained by integrating out both sides of (3.7) w.r.t. the $st$-pair variables
$ds_i d{s_j}' dt_i d{t_j}'$ (giving rise to an $S_3$-symmetric quantity), 
and using (3.9) together with (3.8) as follows:
\begin{equation}  
(2\pi)^3 {\hat G}({\hat \xi} {\hat \eta} ;{\hat \xi}' {\hat \eta}') = 
\sum_{123} D^{-1}({\hat q}_{12}) \int d{{\hat q}_{12}}'' 
K({\hat q}_{12}, {{\hat q}_{12}}'') {\hat G}({\hat \xi}'' {\hat \eta}'';
{\hat \xi}' {\hat \eta}')  
\end{equation}   
This integral equation for ${\hat G}$ which is the 3-body counterpart of
(2.13) for a $qq$ system in the neighbourhood of the bound state pole, 
is the desired 3D BSE for the $qqq$ system in a {\it {fully connected}}
form, i.e., free from delta functions. Now using a spectral decomposition 
for ${\hat G}$ 
\begin{equation}   
{\hat G}({\hat \xi} {\hat \eta};{\hat \xi}' {\hat \eta}')
= \sum_n {\phi}_n( {\hat \xi} {\hat \eta} ;P)
{\phi}_n^*({\hat \xi}' {\hat \eta}';P)/(P^2 + M^2)
\end{equation}   
on both sides of (3.12) and equating the residues near a given pole
$P^2 = -M^2$, gives the desired equation for the 3D wave function $\phi$ 
for the bound state in the connected form:
\begin{equation}   
(2\pi)^3 \phi({\hat \xi} {\hat \eta} ;P) = \sum_{123} D^{-1}({\hat q}_{12})
\int d{{\hat q}_{12}}'' K({\hat q}_{12}, {{\hat q}_{12}}'')
\phi({\hat \xi}'' {\hat \eta}'' ;P)
\end{equation}   
The solution of this equation for the ground state was found in [32] in 
a {\it gaussian} form which implies that $\phi({\hat \xi} {\hat \eta};P)$ 
is an $S_3$-invariant function of ${{\hat \xi}_i}^2 + {{\hat \eta}_i}^2$, 
valid for {\it any} index ``i". While the gaussian form may prove too 
restrictive for more general applications, the mere $S_3$-symmetry of 
$\phi$ in the $({\hat \xi}_i, {\hat \eta}_i)$ pair may prove adequate 
in practice, and hence useful for both the solution of (3.14) {\it and}
for the reconstruction of the 4D BS wave function in terms of the 3D 
wave function (3.14), as is done in Sec.4 below.

\section{Reconstruction of the 4D BS Wave Function}
	To {\it re-express} the 4D $G$-function (3.7) in terms of the 3D 
${\hat G}$-function (3.12), we first adapt the result (2.12) to the hybrid 
Green's function  of the (12) subsystem given by ${\tilde G}_{3 \eta}$, 
eq.(3.10), in which the 3-momenta ${\hat \eta}_3,{{\hat \eta}_3}'$ play a 
parametric role reflecting the spectator status of quark $\# 3$, while the 
{\it active} roles are played by $q_{12}, {q_{12}}' = {\sqrt 3}(\xi_3,
{\xi_3}')/2$, for which the analysis of Sec.2 applies directly. This gives 
\setcounter{equation}{0}   
\renewcommand{\theequation}{4.\arabic{equation}}
\begin{equation}
(2{\pi}i)^2 {\tilde G}_{3 \eta}(\xi_3 {\hat \eta}_3; 
{\xi_3}' {{\hat \eta}_3}') 
= D({\hat q}_{12}){\Delta_1}^{-1}{\Delta_2}^{-1}
{\hat G}({\hat \xi_3} {\hat \eta_3}; {\hat \xi_3}' {\hat \eta_3}')
D({{\hat q}_{12}}'){{\Delta_1}'}^{-1}{{\Delta_2}'}^{-1}
\end{equation}
where on the right hand side, the `hatted' $G$-function has full 
$S_3$-symmetry, although (for purposes of book-keeping) we have not 
shown this fact explicitly by deleting the suffix `3' from its 
arguments. A second relation of this kind may be obtained from (3.7)
by noting that the 3 terms on its right hand side may be expressed in 
terms of the hybrid ${\tilde G}_{3 \xi}$ functions vide their definitions 
(3.11), together with the 2-body interconnection between $(\xi_3,{\xi_3}')$ 
and $({\hat \xi}_3,{{\hat \xi}_3}')$ expressed once again via (4.1), but 
without the `hats' on $\eta_3$ and ${\eta_3}'$. This gives
\begin{eqnarray}
({\sqrt 3} \pi i)^2 G(\xi_3 \eta_3; {\xi_3}'{\eta_3}')
&=& ({\sqrt 3} \pi i)^2 G(\xi \eta; {\xi}'{\eta}')\nonumber\\
&=& \sum_{123} {\Delta_1}^{-1}{\Delta_2}^{-1} (\pi i {\sqrt 3})
\int d{{\hat q}_{12}}'' M d{\sigma_{12}}''
K({\hat q}_{12}, {{\hat q}_{12}}'') 
G({\xi_3}'' {\eta_3}'';{\xi_3}' {\eta_3}')\nonumber\\   
&=& \sum_{123} D({\hat q}_{12}) {\Delta_1}^{-1}{\Delta_2}^{-1}
{\tilde G}_{3 \xi}({\hat \xi}_3  \eta_3; {{\hat \xi}_3}' {{\eta}_3}')
{{\Delta_1}'}^{-1} {{\Delta_2}'}^{-1}  
\end{eqnarray}
where the second form exploits the symmetry between $\xi,\eta$ and 
$\xi',\eta'$. 
\par
	This is as far as we can go with the $qqq$ Green's function, 
using the 2-body techniques of Sec.2. However,, unlike the 2-body case 
where the reconstruction of the 4D $G$-function in terms of the 
corresponding 3D quantity was complete at this stage, the process is
not yet complete for the 3-body case, as eq.(4.2) clearly shows. This
is due to the {\it truncation} of Hilbert space implied in the ansatz 
of 3D support to the pairwise BSE kernel $K$ which, while facilitating a 
4D to 3D BSE reduction without extra charge, does {\it not} have the 
{\it complete} information to permit the {\it reverse} transition 
(3D to 4D) without additional assumptions. This limitation of the 3D 
support ansatz for the BSE kernel affects all $n$-body systems except
$n = 2$ (which may be regarded as a sort of degenerate situation.  
\par
	Now as a purely mathematical problem, we must look for a suitable
 ansatz for the quantity ${\tilde G}_{3 \xi}$ on the right hand side of 
(4.2) in terms of {\it known} quantities, so that the reconstructed 
4D $G$-function satisfies the 3D equation (3.12) exactly, which may 
be regarded as a ``check-point" for the entire exercise. We therefore
seek a structure of the form 
\begin{equation}
{\tilde G}_{3 \xi}({\hat \xi}_3  {\eta}_3; {{\hat \xi}_3}' {{\eta}_3}')
= {\hat G}({{\hat \xi}_3} {\hat \eta}_3; {{\hat \xi}_3}' {{\hat \eta}_3}')
\times F(p_3, {p_3}')    
\end{equation}
where the unknown function $F$ must involve only the momentum of the 
spectator quark $\# 3$. A part of the $\eta_3, {\eta_3}'$ dependence has 
been absorbed in the ${\hat G}$ function on the right, so as to satisfy 
the requirements of $S_3$-symmetry for this 3D quantity, whether it has 
a gaussian structure [32] (where it is explicit), or a more general 
one (where it is not so explicit); see the discussion below eq(3.14).
\par
	As to the remaining factor $F$, it is necessary to choose its
form in a careful manner so as to conform to the conservation of 
4-momentum for the {\it free} propagation of the spectator between two
neighbouring vertices, consistently with the symmetry between $p_3$ 
and ${p_3}'$. A possible choice consistent with these conditions is:
\begin{equation}
F(p_3, {p_3}') = C_3 {\Delta_3}^{-1} {\delta}(\nu_3 - {\nu_3}') 
\end{equation} 
where we have taken only the time-component of the 4-momentum $p_3$ in the 
$\delta$-function since the effect of its space component has already been 
absorbed in the ``connected" (3D) Green's function ${\hat G}$. Next, 
${\Delta_3}^{-1}$ represents the ``free" propagation of quark $\# 3$ between 
successive vertices, while $C_3$ represents some residual effects which may 
at most depend on the 3-momentum ${\hat p}_3$, but must satisfy the main 
constraint that the 3D BSE, eq.(3.12), be {\it explicitly} satisfied.
\par
	To check the self-consistency of the ansatz (4.4), integrate
both sides of (4.2) w.r.t. $ds_3 d{s_3}' dt_3 d{t_3}'$ to recover the 
3D $S_3$-invariant ${\hat G}$-function on the left hand side. Next, in 
the first form on the right hand side, integrate w.r.t. $ds_3 d{s_3}'$ 
on the $G$-function which alone involves these variables. This yields
the quantity ${\tilde G}_{3 \xi}$. At this stage, employ the ansatz 
(4.4) to integrate over $dt_3 d{t_3}'$. Consistency with the 3D BSE, 
eq.(3.12), now demands 
\begin{equation}
C_3 \int \int d\nu_3 d{\nu_3}' {\Delta_3}^{-1} \delta(\nu_3 - {\nu_3}')
= 1 ; (since dt = d\nu) 
\end{equation}
The 1D integration w.r.t. $d\nu_3$ may be evaluated as a contour 
integral over the propagator ${\Delta}^{-1}$ , which gives the pole 
at $\nu_3 = {\hat \omega}_3/M$, (see below for its definition). Evaluating 
the residue then gives 
\begin{equation}
C_3 = i \pi / (M {\hat \omega}_3 ) ;  \quad
{{\hat \omega}_3}^2 = {m_q}^2 + {{\hat p}_3}^2
\end{equation}
which will reproduce the 3D BSE, eq.(3.12), {\it exactly}! Substitution
of (4.4) in the second form of (4.2) finally gives the desired 3-body 
generalization of (2.12) in the form 
\begin{equation}
3 G(\xi \eta; \xi' \eta') = \sum_{123} D({\hat q}_{12}) \Delta_{1F} 
\Delta_{2F} D({{\hat q}_{12}}') {\Delta_{1F}}' {\Delta_{2F}}' 
{\hat G}({\hat \xi_3} {\hat \eta_3}; {\hat \xi_3}' {\hat \eta_3}')
[\Delta_{3F} / (M \pi {\hat \omega}_3)]      
\end{equation}
where for each index, $\Delta_F = - i {\Delta}^{-1}$ is the 
Feynman propagator.
\par
	From this structure of the 4D Green's function near the bound 
state pole, we can infer the corresponding structure of the  4D BS 
{\it {wave function}} $\Phi(\xi \eta; P)$ through a spectral representation 
like (3.13) for the 4D Green's function $G$ on the left hand side of (4.2). 
Equating the residues on both sides gives the desired 4D-3D connection 
between $\Phi$ and $\phi$:
\begin{equation}
\Phi(\xi \eta; P) = \sum_{123} D({\hat q}_{12}){\Delta_1}^{-1}{\Delta_2}^{-1}
\phi ({\hat \xi} {\hat \eta}; P) \times 
\sqrt{{\delta(\nu_3 -{\hat \omega}_3/M)} \over{M {\hat \omega}_3 
{\Delta}_3}} 
\end{equation}
>From (4.8) we can infer the structure of the baryon-$qqq$ vertex function 
by rewriting it in the alternative form [37]:
\begin{equation}
\Phi(\xi \eta; P) = {(V_1 + V_2 + V_3) \over {\Delta_1 \Delta_2 \Delta_3}}
\end{equation}
where 
\begin{equation}
V_3 = D({\hat q}_{12})\phi ({\hat \xi} {\hat \eta}; P) \times
\sqrt{{\Delta}_3 \delta (\nu_3 -{\hat \omega}_3/M) \over{M {\hat \omega}_3}}
\end{equation}
The quantity $V_3$ is the baryon-$qqq$ vertex function corresponding to 
the ``last interaction" in the (12) pair, and so on cyclically. This is 
precisely the form (apart from a constant factor that does not affect
the baryon normalization) that had been anticipated in an earlier
study in a semi-intuitive fashion [37]. 
  
\section{A Simplified 4D NJL-Faddeev Model}

We now set up a simplified 4D NJL-Faddeev bound state problem, 
termed NJL-Faddeev, with 3 scalar-isoscalar quarks interacting 
pairwise in a contact fashion. To facilitate a direct comparison with 
3D-4D-BSE,  we employ the same notation and phase convention for the various
quantities as in Secs.3-4, but in view of the bound state nature of the 
problem it is enough to work with the 4D BSE for the wave function only,
as there is now no special advantage in dealing with the Green's function.
We start with the $qq$ problem as a prerequisite for the solution of the 
$qqq$ problem.

\subsection{$qq$ Bound State in NJL Model}
    
The BSE for the 4D wave function ${\Phi}$ for a $qq$ system may be written
down for the NJL model, in the notation of Sec.2 as :
\setcounter{equation}{0}
\renewcommand{\theequation}{5.\arabic{equation}}
\begin{equation}
i(2\pi)^4 \Phi(q_{12}P_{12}) = {\Delta_1}^{-1} {\Delta_2}^{-1} \lambda
 \int d^4{ q_{12}}' \Phi({q_{12}}'P_{12})
\end{equation}   
where $\lambda$ is the strength of the contact NJL interaction for any
pair of (scalar) quarks. The solution of this equation simply reads as [39]
\begin{equation}
 \Phi(q_{12}P_{12}) = A {\Delta_1}^{-1} {\Delta_2}^{-1}
\end{equation}
When plugged back into (5.1), one gets an  `eigenvalue' equation for the
invariant mass $M_d^2 = -P_{12}^2$ of an isolated bound $qq$ pair in the 
implicit form of a determining equation for $\lambda$: 
\begin{equation}
\lambda^{-1} = -i (2\pi)^{-4} \int d^4q {\Delta_1}^{-1} {\Delta_2}^{-1}
\equiv h(M_d)
\end{equation}
where $\Delta_{1,2}$ are given by (2.2) and $p_{1,2} = P_{12}/2 \pm q$,
viz., $\Delta_{1,2} = m_q^2 + q^2 -M_d^2/4 \pm q.P_{12}$, and we have 
indicated the result of integration by a function $h(M_d)$ of the mass 
$M_d$ of the composite bound state (diquark). Unfortunately the integral 
(5.3) is logarithmically divergent, but it can be regularized with a 4D 
ultraviolet cut-off $\Lambda$, together with a Wick rotation, i.e., 
$q_0 \rightarrow iq_0$, which is allowed by the singularities of the two 
propagators. The exact result is:  
\begin{equation}
16 \pi^2 \lambda^{-1} = 1+ \ln (4\Lambda^2 / M_d^2) -
2 {{\sqrt {4m_q^2-M_d^2}} \over{M_d}} {\arcsin (M_d/{2m_q})}
\end{equation} 
under the condition $M_d < 2m_q$. A slightly less accurate but much simpler  
form which is also easier to adapt to the $qqq$ problem to follow, may be 
obtained by the Feynman method of introducing an auxiliary integration 
variable $u$ ($0 < u < 1$) to combine the two propagators, followed by 
a Wick rotation and a translation to integrate over $d^4q$ (ignoring surface 
terms which formally arise due to the logarithmic divergence) :
\begin{equation}
16 \pi^2 \lambda^{-1} = \ln {6\Lambda^2 \over {6m_q^2 - M_d^2}} - 1 
\equiv 16 \pi^2 h(M_d),  
\end{equation}
thus defining a diquark `self-energy' function $h(M)$ where the `on-shell' 
value is $M=M_d$.  Eq.(5.5) also provides a determining equation for the 
NJL strength parameter $\lambda$ in terms of the `diquark' mass $M_d$ and 
the 4D cut-off parameter $\Lambda$. This result is clearly 4D invariant, 
a feature that characterizes the tenets of the NJL model [17].   

\subsection{NJL-$qqq$ Bound State Problem}

We now set up the corresponding NJL-$qqq$ problem under the same $q-q$
contact interaction strength $\lambda$. Using the same notation for the 
various 4-momenta and propagators as listed  in Sec.3, the 4D wave 
function $\Phi(\xi,\eta;P)$ expressed in terms of any of the $S_3$ 
invariant pairs $(\xi_i,\eta_i)$ of internal 4-momenta satisfies the BSE:
\begin{equation}
i(2\pi)^4 \Phi(\xi,\eta;P) = \sum_{123} \lambda
{\Delta_1}^{-1} {\Delta_2}^{-1} \int d^4q_{12}' \Phi({\xi_3}',{\eta_3};P)
\end{equation}
where the arguments of $\Phi$ on the LHS are not-indexed since it is 
$S_3$-symmetric as a whole, while those on the RHS are indexed in order 
to indicate which subsystem is in pairwise interaction (see explanation 
in Sec.3). The solution of this equation may be read off from the observation 
that the integration w.r.t. $q_{12}'= {\sqrt 3}\xi_3'/2$ leaves the respective 
integrals as functions of $\eta_i$ only, where $i= 1,2,3$. Thus [2]  
\begin{equation}
\Phi(\xi,\eta;P) = \sum_{123}{\Delta_1}^{-1} {\Delta_2}^{-1} F(\eta_3)
\end{equation}
where $F$ is a function of a single variable $\eta_i$. Next, plugging back
the solution (5.7) into the main equation (5.6), gives the following 
integral equation in a single variable $\eta_3$, as a routine procedure 
applicable to separable potentials [2] (see also [34b]) :
\begin{equation}
(h(M_d)- h(M_{12})) F(\eta_3) = -i(2\pi)^{-4} \Delta_3^{-1} 
 \int d^4 q_{12}' [F(\eta_2')\Delta_1'^{-1} + (1 \leftrightarrow 2)]
\end{equation} 
We note in passing that the cut-off parameter $\Lambda$ drops out from the 
LHS, as may be checked by substitution for $h(M)$ from (5.5). This means that 
the 4D diquark propagator  $(h(M_d)-h(M_{12}))^{-1}$ is formally independent 
of the cut-off $\Lambda$, in this simplified NJL model.         
\par
	Next, the meaning of the function $F(\eta)$ can be inferred from
an inspection of eq.(5.8), on similar lines to the 3D [2] or 4D [34b,c]
studies: $F(\eta_3)$ is the 4D `quark(3)-diquark(12)' wave function which
is generated by an exchange force represented by the propagators 
$\Delta_1'^{-1}$ and $\Delta_2'^{-1}$ in the first and second terms on the 
RHS respectively. And the baryon-$qqq$ vertex function $V_3$ corresponding
to a break-up of the baryon into quark(3) and diquark (12) may be identified 
by multiplying this quantity with the product of the inverse propagators 
of quark(3) and diquark(12):
\begin{equation}
V_3 \equiv V(\eta_3) =  \Delta_3 f(\eta_3) F(\eta_3)
\end{equation}     
where the diquark inverse propagator is reexpressed as
\begin{equation}
f(\eta_3) = h(M_d)- h(M_{12}) = (4\pi)^{-2} \ln{{6m_q^2 + \eta_3^2 -4M_B^2/9} 
\over {6m_q^2 -M_d^2}},
\end{equation} 
making use of eq.(5.5) and the kinematical relation 
$\Delta_i = m_q^2+\eta_i^2-M_B^2/9$, where $M_B$ is the mass of the bound 
$qqq$ state, and $i=1,2,3$. The quantity $V_3$ of eq(5.9) may be compared 
directly (except for normalization) with the corresponding `3D-4D-BSE' 
quantity (4.10). 

\subsection{Solution of the Bound $qqq$ State Eq.(5.8)}

We now turn to the Lorentz structure of the NJL-$qqq$ equation (5.8), 
as well as an approximate analytic solution for the energy eigenvalues of
the bound $qqq$ states.  To that end we substitute (5.9) in (5.8) to give an 
integral equation for $V(\eta_3)$, with $\eta_2' \equiv \eta$ for short: 
\begin{equation}
V(\eta_3) = -2i(2\pi)^{-4} \int d^4\eta V(\eta) f^{-1}(\eta) \times 
(m_q^2 +\eta^2 - M_B^2/9)^{-1} (m_q^2 + (\eta_3 + \eta)^2 - M_B^2/9)^{-1}
\end{equation}
where the factor $2$ on the RHS signifies equal effects of the two terms
on the RHS of (5.8). For a bound state solution of this equation, with
$M_B < M_d + m_q$, the singularity structures permit a Wick rotation 
$\eta_0 \rightarrow i\eta_0$ which converts $\eta$ into a Euclidean variable
$\eta_E$. This shows without further ado that eq.(5.11) is 4D-invariant
just like its $qq$ counterpart eq.(5.3). This is not quite the same thing as
the old result [13] on O(4)-like spectra with harmonic confinement in the
limit of infinite quark mass [13], since this NJL-Faddeev model of contact
interaction, patterned after similar approaches [34], lacks a confining 
interaction, so that although in principle eq.(5.11) predicts a spectrum 
of bound states at the $qqq$ level (starting with  NJL(contact) pairwise 
interactions), such spectra cannot be a realistic representation of the 
{\it actual} hadron spectra [12]. (Note that the 4D form factors employed 
in some NJL interactions [34b] for convergence of the integrals, do not have 
the significance of confinement; see however other attempts at confinement
[40]). In the absence of a confining mechanism, most such NJL-Faddeev models 
[34] have effectively produced only one non-trivial bound state - the 
nucleon/Delta. We now show how this comes about via Wick rotation in (5.11).     
\par
	For an approximate analytic solution of eq.(5.11), note that the 
logarithmic function $f(\eta)$ in the integral appearing on the right is 
slowly varying, so that not much error is incurred by taking it out of the 
integral and replacing it with an average value $<f(\eta)>$. It is now 
possible to  `match' both sides with an  effectively constant $V(\eta)$, 
{\it provided} any further logarithmic dependence on $\eta$ is also similarly 
treated for consistency. The integral is now exactly of the type (5.3), i.e., 
logarithmically divergent,  and can be handled successively by Wick rotation, 
Feynman auxiliary variable $u$, and a translation. The result is again of the 
form (5.5), and after cancelling out the factors $V(\eta_3)$ and $V(\eta)$ 
from both sides, the eigenvalue equation reads: 
\begin{equation}
<f(\eta)> = 2 (4\pi)^{-2} [\ln {\Lambda^2 \over 
{<\eta_3^2/6> + m_q^2 -M_B^2/27}} - 1]  	
\end{equation}
To simplify this equation, we express all quantities in terms of the $h(M)$
functions given in (5.5) and (5.10), and ignore the difference between 
$\eta = \eta_2'$ and $\eta_3$ inside the logarithms, to give
\begin{equation}
h(M_d)-h(M_{12}) = 2h(M_{12}); \Rightarrow \lambda^{-1} = h(M_d) = 3h(M_{12}) 
\end{equation}
The last equation brings out clearly the fact that the baryon binding comes
about from {\it three} pairs of $qq$ interaction, albeit off-shell, since
the function $<M_{12}^2> = <\eta_3^2> -4M_B^2/9$ still depends on the 
(average) value of $\eta^2$. The qualitative features are thus on expected 
lines, but this oversimplified model was not intended to be pushed for an 
actual fit to the nucleon mass (which at minimum requires the introduction of 
spin-isospin d.o.f.[34]), beyond the general feature of a quark-diquark
structure that characterizes the NJL-Faddeev approach [34], as expected from 
any separable potential [2], of which the NJL model is a special case. The
`bosonization' approaches [36] also bring out the quark-diquark structure,
but in a more general (non-linear) fashion [36b]. 

\subsection{Comparison of NJL-Faddeev with 3D-4D-BSE}     

We end this section with a comparison between the vertex functions (4.10) of 
3D-4D-BSE [33], and (5.9) of NJL-Faddeev [34], which reflect the corresponding 
differences in their respective dynamical premises. The NJL-Faddeev form 
(5.9) of $V_3$ is explicitly Lorentz invariant, with full 4D Hilbert space 
information built into its structure. Its  quark-diquark structure merely 
reflects the `separable' nature of the NJL model [17]. There is no special 
motivation here for a 3D reduction of the types [5-8] described in Sec.1. 
The 4D (Lorentz-invariant) structure is in-built in NJL-Faddeev [34], as seen 
from eqs(5.4-5) for the $qq$ state, and its $qqq$ counterpart in eq.(5.8).  
\par
	In contrast, the vertex function (4.10)  obtained from the 3D-4D-BSE 
formalism [9-11] is merely Lorentz covariant due to the 3D kernel support, 
but the derivation is otherwise quite general (much more than NJL-Faddeev)
since it is valid for {\it any} form of the kernel as long as it is 3D in 
content. This leads to an exact 3D reduction of the (4D) BSE  whose formal 
solution is a 3D wave function $\phi({\hat \xi},{\hat \eta})$, a function 
of {\it two} independent 3-momenta [32], in contrast to its NJL counterpart 
$F(\eta_3)$ in (5.9) which is a function of a single 4-momentum $\eta_3$. 
The denominator function $D({\hat q}_{12})$ of (4.10) similarly is a 3D 
counterpart of the corresponding 4D inverse propagator $f(\eta_3)$ in (5.9). 
Finally the big radical in (4.10) corresponds to the inverse propagator 
$\Delta_3$ in (5.9), but has a vastly more involved structure which may be 
traced back to the difference in their respective dynamical premises which 
we now seek to account for analytically.  
\par
	While the premises of NJL-Faddeev [34] are traceable to the contact
4-fermion interaction [17], those of 3D-4D-BSE [9-11] are based on an 
interplay of the 3D and 4D BSE forms with an otherwise general but 3D 
kernel support. To  reconcile the two approaches in a mathematically 
consistent way, note first that while the `zero extention' in the temporal 
direction is common to both, NJL-Faddeev has also a zero spatial extension, 
but 3D-4D-BSE has `normal' spatial extension. To reduce the latter to the 
former, all that is required is to set its spatial extension to zero ! This is 
simply achieved (in momentum space) by setting its 3D kernel $K$ equal to a 
{\it constant}. For, as is easily checked in this limit, the structures of 
the various {\it 4D} equations in Secs.2-4 reduce exactly to their 4D NJL 
counterparts in Sec.5.(1-3), via the simple identification $K = \lambda$, 
so that there is no more need to make 3D reductions or 4D reconstructions! 
[The conversion from the Green's function form of Secs.2-4 to the wave 
function language of secs.5.(1-3) is trivial]. Indeed the formulation of 
NJL-Faddeev in this Section has been made in close enough correspondence 
with that of 3D-4D-BSE to bring out the transparency of the parallelism.  
\par
	After this basic check between these two formalisms, it is clear
that the ambiguity in the reconstruction of the 4D wave function from the 
3D form vanishes in the $K = const$ limit, so that the same is directly 
attributable to the (mere Lorentz covariant) 3D form of the BSE kernel. 
Indeed from the derivation in Sec.4 it is clear that the 1D $\delta$-function 
in (4.10) fills up an information gap in the reconstruction from a truncated 
3D to the full 4D Hilbert space in the simplest possible manner, while 
satisfying a vital self-consistency check by reproducing the full structure 
(3.12) of the 3D BSE. This already lends {\it sufficiency} to the ansatz 
(4.4) which leads to (4.10). As to its `necessity', this ansatz has certain 
desirable properties like on-shell propagation of the spectator in between 
two successive interactions, as well as an explicit symmetry in the $p_3$ 
and ${p_3}'$ momenta. There is a fair chance of its uniqueness within some
general constraints, but so far we have not been able to prove this.  
\par
	The other question concerns the compatibility of the 1D $\delta$-
function in (4.10) with the standard requirement of connectedness [3-4]. 
Both the $\delta$-function and the $\Delta_{3F}$ propagator appear in 
{\it rational} forms in the 4D Green's function, eq.(4.7), reflecting 
a free on-shell propagation of the spectator between two vertex points.
The square root feature in the baryon-$qqq$ vertex function (4.10) is a
technical artefact resulting from equal distribution of this singulariity 
between the initial and final state vertex points, and has no deeper 
significance. Furthermore, as the steps in Sec.3 indicate, the three-body 
connectedness has already been achieved at the 3D level of reduction, so 
the `physics' of this singularity, generated via eq.(4.4), must be traced 
to some mechanism other than a lack of connectedness [3,4] in the 3-body 
scattering amplitude. A plausible analogy is to a sort of (Fermi-like) 
`pseudopotential' of the type employed to simulate the effect of chemical
binding in the coherent scattering of neutrons from a hydrogen molecule 
in connection with the determination of the {\it singlet} $n-p$ scattering 
length [38]. Such $\delta$-function potentials have no deeper significance 
other than depicting the vast mismatch in the frequency scales of nuclear
and molecular interactions. In the present case, the instantaniety in time 
of the pairwise interaction kernel in an otherwise 4D Hilbert space causes a 
similar mismatch, needing a 1D $]delta$-function to fill the gap. And 
just as the `pseudo-potential' in the above example [38] does not have
any observable effect, the singularity under radicals in (4.10) will 
{\it not} show up in any physical amplitude for hadronic transitions via 
quark loops, since the Green's functions (4.7) involve both the $\delta$-
function and the propagator $\Delta_{3F}$ in {\it rational} forms before 
the relevant quark loop integrations over them are performed.
\par
	As to the wider ramifications on spectroscopy, the $L$-excited $qqq$ 
states that characterize the hadron spectrum, are effectively absent in the  
NJL-Faddeev eq.(5.11), unlike in 3D-4D-BSE [31-32] where they do. On the 
other hand, both vertex functions (4.10) and (5.9) are perfectly capable of 
generating amplitudes for baryonic transitions via quark-diquark loop 
integrals, as ahown with NJL-Faddeev [41] for  items such as axial coupling 
constants, nucleon magnetic moments and the $\pi-N$ $\sigma$-term, and with
3D-4D-BSE for similar items [42,43].
  
\section{Discussion and Summary}

We first recapitulate the perspective on 3D reduction of the BSE, outlined 
in sec.(1.2): It involves both  conceptual (simultaneity of constituents in 
a bound state) and observational (O(3)-like spectra [12]) issues. In this 
respect, the traditional 3D approaches to the BSE [5-8] work with normal
4D kernel support but manipulate the associated propagators in the 4D BSE in 
various ways [6-8] to reduce it to a 3D form. There is no formal problem of
compatibility with the observed O(3) spectra [12] in such $qqq$ approaches, 
except that, to the author's knowledge, there is little evidence so far of 
such methods going beyond the 2-quark level. Anyhow, in such methods, the 
reduced 3D BSE represents a fresh starting point of the dynamics, and there 
is no going back to its earlier 4D form. Because of this reason, one does not
see in these approaches [6-8] such unorthodox radicals as in (4.7-10) which 
characterizes the alternative approach [9,10] based on the Markov-Yukawa 
Transversality Condition [11]: It postulates a 3D support to the BSE kernel 
in a covariant fashion by demanding the internal 4-momenta to be transverse 
to the total 4-momentum of the composite hadron, and allows an {\it exact} 3D 
reduction of the original 4D BSE [9-10], thus automatically ensuring 
O(3)-spectra. More importantly, it allows an {\it exact} reconstruction of the
4D BSE amplitude without extra charge, so that both BSE forms are completely 
equivalent and therefore simultaneously available for both 3D spectroscopy 
and 4D quark-loop amplitudes in a `two-tier' fashion [14]. (This was shown 
for the 2-quark problem some years ago [9], and the present exercise [33] 
is for the corresponding 3-quark problem).

\subsection{Comparison with Other 4D Approaches: Spectroscopy}
   	
The principal results of this paper on the structure of the 3D-4D-BSE 
formalism [9,33,37] for the $qqq$ problem are contained in eqs(4.7-10).
For comparison with other contemporary approaches to the $qqq$ problem, we 
have chosen for explicit display a simplified form of the 4D NJL-Faddeev 
model [34], which nevertheless retains its principal features, and outlined 
in Sec.5 the derivation of its main results through a formulation designed 
to bring out the parallelism with 3D-4D-BSE of Secs.(2-4) in point-wise 
details. In particular, the equations of 3D-4D-BSE {\it match} exactly those 
of NJL-Faddeev by setting the BSE kernel equal to a constant. This calibration
ensures that any ambiguity inherent in the reconstruction of the 4D baryon-
$qqq$ vertex function in Sec.4 fully disappears in this zero range limit, so 
that the difference between these 2 approaches may be directly attributed to 
the finite spatial range of the 3D-4D-BSE formulation. The respective vertex 
functions also stand a close comparison, except for the 2-body (quark-diquark) 
structure of NJL-Faddeev (due to the separable interaction [2]), versus the
genuine 3-body structure of 3D-4D BSE. The only item of dissimilarity 
is in respect of the singularity factor under radicals in (4.10), vis-a-vis 
the inverse propagator $\Delta_3$ of the spectator in (5.9). This has been 
accounted for in sec.(5.4), mathematically through self-consistency checks,
and physically by an analogy with psudopotentials of the type employed for 
coherent scattering of slow neutrons from hydrogen molecules [38], with no 
observable effects. In particular, the 1D $\delta$-function in no way implies 
a lack of connectedness [3-4] in the $qqq$ amplitude.         
\par
	As regards spin, the extension of the above formalism to fermion 
quarks is a straightforward process amounting to the replacement of 
$\Delta_F = -i \Delta^{-1}$ by the corresponding $S_F$-functions, as 
has been described elsewhere [42,37]. In particular, the fermion vertex 
function has recently been applied to the problem of proton-neutron mass 
difference [43] via quark loop integrals, to bring out the practicability 
of its application without parametric uncertainties, since the entire 
formalism is linked all the way from spectroscopy to hadronic transition 
amplitudes [14] via quark-loops.
\par
	In sec.(1.2) we have attempted briefly to set this 3D-4D BSE-SDE 
approach in the physical context of other 4D QCD-motivated BSE-SDE approaches 
[21,16] which are all governed by $DB{\chi}S$ on the one hand, and the 
large $N_C$ limit on the other. These are more general than the contact NJL 
model [17,34] since, unlike in [17,34], their non-local kernel structures 
allow incorporation of confinement for prediction of $L$-excited hadrons. 
Our {\it vector}-type (gluon-like) kernel [30] for both the perturbative and 
non-perturbative QCD regimes gives it a chirally invariant look at the 
4-quark Lagrangian level, and with the constituent mass $m_q$ `understood' 
via $DB{\chi}S$ [15], fits in quite well within a broader BSE-SDE scenario 
[21,16], except for its hybrid 3D-4D content [9]. 
\par
	This last brings us to the spectroscopy implications of 3D-4D-BSE 
vis-a-vis  other regular 4D BSE models [21,16,34,36]. Thus NJL-Faddeev [34] 
supports one non-trivial bound state - the nucleon (see Sec.5), in the absence
of a confinement mechanism, although a separable $qq$ interaction {\it per se}
does not rule out excited states at the 3-body level. The solitonic baryon 
models [29,35] also have similar predictions. In contrast, 4D BSE's [16] with 
their {\it local} 4D interaction kernels, have in principle the capacity to 
predict $L$-excited states as well, except that their actual applications to 
$q{\bar q}$ mesons have so far not gone beyond the ground $L = 0$ states, 
albeit for many flavour/spin combinations [16], while the $qqq$ spectroscopy 
of such BSE-SDE models [16]  is not yet in sight. In this regard, the 
predictions of such models [16] on genuinely $L$-excited spectra would be 
of considerable interest for resolving the old issue of O(4)-like spectra 
found within a Wick-rotated BSE formalism under harmonic confinement [13].          
\par
	In contrast, the 3D-4D-BSE formalism [9,33] seems capable of 
encompassing both 3D spectroscopy [31-32] and 4D matrix elements within
a common dynamical framework. The only cost factor is the information gap 
in the reconstruction of the 4D vertex in terms of 3D ingredients; the 1D
$\delta$-function in the baryon-$qqq$ vertex function, eq.(4.10), fills up
the gap, and is accounted for in detail in sec.(5.4). Another positive 
aspect of the 3D-4D formalism lies in the generality and flexibility of its 
3D kernel structure, since the 3D-4D interconnection [9,33] does {\it not} 
depend on a precise knowledge of its functional form. And with its dynamics 
already rooted in spectroscopy [31-32], it has so far given parameter-free 
predictions on magnetic moments [42] as well as the $n-p$ mass difference 
[43], the latter being 1.28 MeV (vs. expt 1.29).            
\par              
	Yet another type of approach to the $qqq$ problem, as available 
in the literature, concerns parametric representations attuned to 
QCD-sum rules [44], effective Lagrangians for hadronic transitions to
``constituent" quarks, with ad hoc assumptions on the hadron-$qqq$ form 
factor [45], similar (parametric) ansatze for the hadron- quark-diquark 
form factor [46]; or more often simply direct gaussian parametrizations 
for the $qqq$ wave functions as the starting point of the investigation 
[47]. Such approaches are often quite effective for the investigations of
some well-defined sectors of hadron physics with quark degrees of freedom,
but are in general much less predictive than dynamical-equation-based 
methods like NJL- Faddeev [34] or 3D-4D-BSE [33], when extended beyond their 
immediate domains of applicability. 

\subsection{Summary and Conclusion}

To summarise, this work arose out of the need for a formal demonstration 
[33] of a semi-intuitive ansatz [37]  for the reconstruction of the 4D  
baryon-$qqq$ vertex function in terms of 3D ingredients, as in an earlier 
work for the 2-quark case case [9], under conditions of a covariant 3D 
support to the pairwise BS kernel. This has been viewed as a self-contained
mathematical problem in its own right for an otherwise general 3D kernel,
in the hope that the rich potential of such approaches as the 3D-4D-BSE 
formalism to access both spectroscopy and quark-loop amplitudes, under a 
common umbrella for 2- and 3-quark hadrons,  will prove useful to others 
wishing to explore similar variants of the Transversality Principle [11].
The unfinished task in the 3D-4D programme has now been completed via Green's 
function techniques for both the 2- and 3- body problems, with a prior 
calibration to the 2-body case. In the process, the loss of Hilbert space 
information involved in the 4D reconstruction has been made up with 
the help of certain 1D $\delta$-functions signifying on-shell propagation 
of the spectator $\# 3$ between two successive interactions of the 
$12$-pair, but which in no way violate connectivity [3,4]. The central 
result, eq.(4.10), for the reconstructed 4D vertex function [33], has 
been fully accounted for through a detailed comparison with the structure
of the NJL-Faddeev model [34].   
\par
	The entire approach in the programme has been conceived with due
emphasis on the need to include the spectroscopy sector as an integral part
of any `dynamical equation based' approach. This perspective is hardly new, 
having been set more than 25 years ago by none other than Feynman et al [48], 
but can certainly stand a reiteration.   
\par
	The initial version of this work [33]  was written at the 
International Centre for Theoretical Physics during a short visit by the
author in 1996 end. This hospitality of the Director, Prof.M.A.Virasoro, 
is gratefully acknowledged. The author is grateful to some critics of [33] 
for bringing to his notice related approaches like NJL-Faddeev to his 
attention and suggesting a critical comparison with them, which led to the
present version. He also appreciates help from Dr. Usha Kulshreshtha in
finding some key references on NJL-Faddeev models.

\end{document}